\documentstyle[12pt]{article}
\begin{document}

\begin{center}

{\bf The "phenomenological" space time and quantization.}

\vspace{1.3cm}
Marcelo BOTTA CANTCHEFF\footnote{e-mail: botta@fis.uncor.edu} \\
\vspace{5mm}
FaMAF, Medina Allende y Haya de la Torre,\\
Ciudad Universitaria, 5000 C\'ordoba, Argentina\\
 Fax nr: +54-51-334054\\
{\it Personal address:} General Paz 540 3-D, \\
5000 Cordoba, Argentina.

\end{center}

\newcommand{\be}{\begin{equation}}
\newcommand{\ee}{\end{equation}}
\newcommand{\epe}{\end{equation}}
\newcommand{\bea}{\begin{eqnarray}}
\newcommand{\eea}{\end{eqnarray}}
\newcommand{\ba}{\begin{eqnarray*}}
\newcommand{\ea}{\end{eqnarray*}}
\newcommand{\epa}{\end{eqnarray*}}
\newcommand{\ar}{\rightarrow}

\vspace{5cm}

\begin{abstract}

  In order to obtain a well defined quantum gravity we define the spacetime in relation to the "phenomenology" of the physical interactions. Under certain conditions the Feynman Rules may be the literal description of this phenomenology; however we shall to speculate with this {\it in general}. Besides, we comment the reasons that give to the gravitational field a privileged situation over the others.
\end{abstract}
\vspace{5cm}
 The usual meaning of spacetime will be reviewed. The idea is simple: in general, only certain subset of the current spacetime is involved in the evolution of a given physical system (we are interested in quantum systems) or observations; only this has effective physical significance and its description is required. furthermore, to consider the rest of the events can to introduce difficulties unnecessarily, in particular, in order to obtain a well defined quantum gravity. Each "effective" event must be associated with a "transformation" (or observation) of a physical system. The spacetime (the set of the events) would be not given before the fields, it develops together with the interactions and obviously, the causal structure of the spacetime is in correspondence with the causal connections between the elemental phenomena that compound the evolution of the full system. We call to this concept: "Phenomenological Space Time (P.S.T.)". Notice that naturally, within this context {\it the space time need not to be continuous} \cite{bomb}.
\\
 The mechanism proposed is: "if some full dynamic scheme for the fields is assumed, the P.S.T. {\it relative to a particular process}, will be the set of support-points on which it developes, in consistence with that dynamic."
\\
 
This concept could be successfully applied in the quantization only if a "modification" of the usual dynamic for the phenomena is assumed; the canonical formulation of quantum field theory \cite{1} \footnote{ That is to say, Euler-Lagrange equations on a continuous spacetime.} requires a continuum background and is not clear the meaning  of an event in the sense P.S.T.\\
Thus, the question is : what does the phenomenology mean, in general?; what is, precisely a transformation of a system in the framework of quantum field theory? We attempt to introduce the answer.
\\
For this, it is convenient to formulate the dynamic in terms of an general expression for the transition amplitude between two quantum states, which in general will be some functional of the space time points involved in the process where the fields develope; for example the sum over histories. Then we shall to define the $M$-spacetime such that the transition amplitude let "well behaved". It signify to allow only a spacetime {\it such that it avoid the singularities} and so, the problems of infiniteness of quantum gravity would be solved {\it by construction} \footnote{ Which constitues a substantial motivation for this review of the current concepts.}\\
 Now, we illustrate these ideas assuming as {\it generic} the "scenery" where the perturbation theory is built :
\\
  In {\it The Covariant Perturbation Method}(C.P.M.)\cite{2}, the gravity is treated \footnote{The expansion parameter in the perturbation series is the squared Plank length, $l_p^2$.} as a spin-2 field $\gamma_{ab}$ on a background \footnote{most generally, it could be a solution $(M_0 ; g^0_{ab})$ of the Einstein's equation} $ (R^4,\eta_{ab}=diag(-1;1;1;1))$. The full metric is $ g_{ab} \equiv  \eta_{ab} + \gamma_{ab} .$
The Lagrangian for gravity coupled with the "matter fields" is obtained replacing $\eta_{ab} \to g_{ab}$ and $\partial_a \to \nabla_a$\footnote{ the canonical connection.}.\\
where $\gamma_{ab}$ is not assumed be small in a set $\Omega$ inclouded in a compact subset of $R^4$ and $g_{ab}$ satisfy the exact Einstein's equation; in the spacetime complement\footnote{the spacetime is the union of $\Omega_0$ and $\Omega$} $\Omega_0$, $\gamma_{ab}$ is small and the the interactions are considered negligible. \\
The commutation relations and the causality conditions are defined with respect to $\eta_{ab}$. Notice that the role of $\eta_{ab}$ is analogous to an external electromagnetic field $A^a_{ext}$ in Q.E.D.; it must be treated classically.
\\
Then formally, one can to construct the Feynman's Rules (F.R.) in the configuration space. The F.R., illustrated by the Feynman's Diagrams, are constructed as a "pictorial" description of the mathematics involved in the perturbation schemme; they allow an interpretation of the fundamental interactions strongly intuitive, and it lead us to conjeture that in this context, they must not be interpreted only as a formal illustration and {\it gives a precise meaning to the "phenomenology" mentioned above}
\\
In order to apply the mechanism of suggested by the P.S.T and
the complete agreement of the predictions of perturbative scheme with the existing experiments allows us to assume\footnote{ of course,{\it under the conditions stablished above}.}:
\\
{\bf (\#)} "the F.R. (in the Configuration Space\footnote{ perhaps it is a unnatural restriction in {\bf (\#)}, because the construction of F.R. in momentum space is not direct,they assume the momentum conservation in each vertex; but if $\Omega$ is discrete the Noether's theorem cannot be applied because the notion of the {\it operations of symmetry} on the spacetime changes.}) {\it are not a mathematical artificer}, they represent the fundamental structure of the processes ".
\\
 So, the canonical approach "will be an {\it approximation}" which can be recovered in the "classical" limit, when the quantum effects of the gravity are negligibles.
\\ 
 In this scheme a "transition" (virtual) between two quantum states is described pictorially, and now {\it literally}, with a vertex of a Feynman Diagram; then "for each space time point there exist a set of such vertex associated to the full process occurring in whole the spacetime."
\\
  This is the new precise structure of a spacetime event in terms of the physical processes such as we thought.
\\
 In each order, the $S$-matrix expansion is expressed \cite{1} by: 
\be
S^{(n)}[\Omega] \equiv \frac{i^n}{(n!)}\int_{\Omega} dx_1..\int_{\Omega} dx_n T[{\cal L}_I(x_1)..{\cal L}_I(x_n)] , \label{s}
\ee
where ${\cal L}_I$ is the interaction lagrangian density whose support is $\Omega$\footnote{$T$ is the time ordered product}. 
\\
  When $\Omega$ contains a finite number of elements, the integral is replaced by a sum over its elements and the lagrangian density must be interpreted as a distribution. Formally, the partial derivatives are taking with respect to the background coordinates.
\\
  {\bf (\#)}, not only does connect the dynamic for the topological structure of the spacetime with the evolution of the fields, besides it has the necessary kinematic flexibility to avoid the known infiniteness of perturbation theory. Only will be allowed a spacetime, such that it does avoid those infiniteness.
\\
 That is expressed in the following definition;\\
 {\bf (*)} Given a Fock-Space ${\cal F}$ (built with the solutions of free fields), and a "initial" state $|\psi_{in}> \in {\cal F}$, then:"for a quantum transition ($|\psi_{in}> \to |\psi_{out}> \in {\cal F}$); M is a {\it subset} of $R^4$ such that, each corresponding element of the $S$-matrix expansion \cite{1} remains finite in each order"; and the vector $S|\psi_{in}>$ have norm unity.
\\
 C.P.M. has two high difficulties; first, the infiniteness in the perturbation theory for gravity; we sketch to show that "there exists {\it at least} a trivial spacetime $M$ such that the $S$-matrix expansion is finite for all quantum process": \\
the {\it minimal} spacetime $M=\Omega_0$ where the $S$-matrix is trivially well defined: $<\psi_{out}|S[\Phi]|\psi_{in}> = <\psi_{out}|\psi_{in}>$, and the interaction region $\Omega_0$, would be the empty set $\Phi$.\\
Thus, in principle, we could "to enlarge $\Omega_0$" {\it adding} points such that $S$ remains finite. 
 \\
 Clearly, two spacetimes will be {\it equivalents} when they have equal distribution of probabilities given by the $S$-matrix .\\
Note that, in order to define more precisely $M$; we have adopted the "extra" restriction: "$M$, is some subset of $R^4$". Also we could to assume that
 $M$ is {\it maximal} in the following sense; that it cannot be included into any other $M'$ (both included in $R^4$) satisfying {\bf (*)}". Which ensures that when the quantization of the gravity is not considered and the theory is renormalizable; the perturbation series is finite in each order thus, we don't need to constrain the set of events to avoid the infinities. {\it $M$ agrees with the continuum background in consistence with the canonical field theory which do not includes gravity} (for example Q.E.D. on Minkowski space time). However it can be not sufficient to determinate completly the spacetime $M$\footnote{The uniqueness of $M$ will be not analyzed here; the equivalence of spacetimes must be studied with detail enough\cite{botta}.}; the inclusion defines a "partial order" and each class totally ordered may to have an maximal element\footnote{ $M$ is upper bounded by $R^4$, thus, it is clear that it there exist as the limit point of a totally ordered (with respect to the inclusion) succession of sets $M_N$ (for which the $S$-matrix are defined); unless that such succession converges to a set $M'$, whose scattering matrix diverges. In this case we define the maximal element as {\it the union of all the $M_N$}; for this, however, is not clear how to evaluate the $S$- matrix.} \\
Other alternatives can be proposed; for example, perhaps one could to define other scheme with a natural probability for each spacetime which satisfies {\bf (*)}, in this case we do not need to determinate $M$ univocally.\\
  The second problem in C.P.M. is about the ambiguity in the causal structure of the space time; the causality conditions holds with respect to the background metric $\eta_{ab}$ rather than the true $g_{ab}$ but in the free region $\Omega_0$,  $g_{ab}$ is accurately $\eta_{ab}$, and {\it the causal structure} of both metric agree. Besides we hope that in general, $\Omega$ contains a finite number of elements \footnote {This is a conjecture; the proof could be that if any subset of $\Omega$ is a continuous the infinities goes back to appear.} when the gravity is considered; if this is true: {\it both causal structure agree in all space time region} in elaborate but nonetheless well posed sense: the notion of the causal structure is based on the type of the curves; let a timelike curve $\gamma$ on the background, then $\gamma$   intersects to $\Omega$ in a discrete set of points thus, the proper time of $\gamma$ do not changes when the true metric is considered. It is to say: "the metric $g_{ab}$ in $\Omega$ (discrete) is ill defined; however 
is possible to define the causal structure with respect to the background $(R^4, \eta_{ab})$ without ambiguity".
\\
  Notice, on the other hand, that there is an {\it algebraic explanation} of the privileged situation of the metric field over the others: {\it the quantized gravitational field determines the the spacetime points} as consecuence of {\bf (\#)}:\\
   The expression (\ref{s}) requires a metric $g_{ab}$ in each point $x$ to "contract" the tensorial (or spinorial \footnote{ The metric field can be described with spinors.}) indexes of the fields interacting \footnote{For example in Q.E.D. the Interaction Lagrangian is: ${ \cal L }_I (x) = -A_{\mu}(x)s^{\mu}(x) = -A^{\mu}(x)s^{\nu}(x)g_{\mu \nu}(x)$ the metric is needed to make the contraction between the indexes $\mu,\nu$.}. Thus $g_{ab}$ must be present in each term of the Langrangian, it to say that the metric $g_{ab}$ "necessarily" interacts with all the other fields in each spacetime point. {\it Each one of the "events" is a vertex with at least one "gravitonic line" on it.}
\\
 This appears as extremly trivial way to avoid the problem of non renormalizability of C.P.M. however; a assumption like {\bf (\#)} does justify it!\\
 The most significant aspect of this scenery is to fix the attention on the evolution of the fields which "develops" the spacetime. This is a {\it ontologic} requirement, believed only a philosophical question without relevance in physics.\\
 Surely there is no an unique way to develop this.\\
\\
Finally,\\
\\
 the P.S.T. depends critically of the "phenomenology" assumed, later
 we speculate that:\\
 "the F.R. constitue the {\it fundamental entity}", and {\bf (\#)} is the fundamental (and strongly intuitive) description of {\it all} quantum process" not only in the conditions of perturbation theory.\\
 Clearly, this could does affect more strongly the spacetime structure and the physical laws.\\

\end{document}